\documentclass[useAMS,usenatbib]{mn2e}
\pdfoutput=1
\usepackage{graphicx}
\usepackage{dcolumn}
\usepackage{bm}
\usepackage{amssymb,amsmath,bm}
\usepackage{color}
\usepackage[colorlinks,linkcolor=red,citecolor=blue,urlcolor=blue ]{hyperref}
\usepackage{multirow}
\usepackage[utf8]{inputenc}
\usepackage{balance}
\usepackage{enumitem}
\usepackage{lipsum}
\usepackage{ulem}

\title[Investigating the morphology of SN 1006]{Numerically investigating the morphology of the supernova remnant SN 1006 in the ambient medium with a density discontinuity}
\author[Jun Fang et al.]{Jun Fang$^1$, Jingwen Yan$^1$, Lu Wen$^1$, Chunyan, Lu$^1$, Huan Yu$^2$\\
                      $^{1}$Department of Astronomy, Key Laboratory of Astroparticle Physics of Yunnan Province, Yunnan University, Kunming 650091, China; \\fangjun@ynu.edu.cn\\
                      $^{2}$Department of Physical Science and Technology, Kunming University, Kunming 650214, China; yuhuan.0723@163.com
                      }

\begin{document}
  \date{\today}
  \pagerange{1--18} \pubyear{2019}
  \maketitle

\begin{abstract}
  Multiband observations on the type Ia supernova remnant SN 1006 indicate peculiar properties in its morphologies of emission in the radio, optical and X-ray bands. In the hard X-rays, the remnant is bilateral with two opposite bright limbs with prominent protrusions. Moreover, a filament has been detected at the radio, optical and  soft X-ray wavelengthes. The reason for these peculiar features in the morphologies of the remnant is investigated using 3D HD simulations. With the assumption that the supernova ejecta is evolved in the ambient medium with a density discontinuity, the radius of the remnant's boundary is smaller in the tenuous medium, and the shell consists of two hemispheres with different radiuses. Along particular line of sights, protrusions appear on the periphery of the remnants since the emission from the edge of the hemisphere with a larger radius is located outside that from the shell of the small hemisphere. Furthermore, the northwest filament of SN 1006 arises as a result of the intersection of the line of sight and the shocked material near the edges of the two hemispheres. It can be concluded that the features that the protrusions on the northeast and southwest limbs and the northwest filament in the morphologies of SN 1006  can be reproduced as the remnants interacting with the medium with a density discontinuity.
\end{abstract}

\begin{keywords}
  Hydrodynamics (HD) -- methods: numerical -- ISM: supernova remnants
\end{keywords}

\section{Introduction}\label{sec:intro}

Supernova remnants (SNRs) are important candidates of the sites of the cosmic rays with energies up to $\sim 3\times10^{15}\, \mathrm{eV}$ accelerated in the Galaxy, and many of them have been identified at radio, X-ray and $\gamma$-ray wavelengthes. For example, SN 1006, which relates with the type Ia supernova in AD 1006 from Chinese records \citep{2010A&G....51e..27S}, is a well-known SNR widely detected in the radio, optical, X-ray and $\gamma$-ray bands \citep{1997ApJ...491..829W,2003ApJ...589..827B,2009AJ....137.2956D,2017ApJ...851..100C,2018ApJ...864...85L,2019PASJ...71...77X}, and it  is located at a distance of $2.2\,\mathrm{kpc}$, $14.6^{\circ}$ above the Galactic plane around by a relatively low-density interstellar medium \citep{2003ApJ...585..324W}. The spectrum of the nonthermal emission from radio to X-rays can be well reproduced as the synchrotron radiation from a population of high-energy electrons with an energy distribution of a power law plus an exponential cutoff \citep{2018ApJ...864...85L}.

Interesting morphologies of emission for SN 1006 have been obtained through multiband observations. In the hard X-rays, two bright  limbs are located in the northeast (NE) and southwest (SW) of the remnant from the images obtained with ASCA \citep{1995Natur.378..255K}, XMM-Newton \citep{2015MNRAS.453.3953L}, Chandra \citep{2003ApJ...589..827B,2014ApJ...781...65W}, and NuSTAR \citep{2018ApJ...864...85L}. The feature of two bright lobes also exists at the radio wavelength \citep{1986AJ.....92.1138R,2003ApJ...586.1162L}. The two bright limbs in the band of radio to X-rays are power-law dominated as a result of the synchrotron radiation of the relativistic electrons accelerated by the shock. Especially, there are several thin filaments outlining the edges of the shell in the NE and SW, and they are sometimes crossing each other \citep{2003ApJ...589..827B,2004A&A...425..121R,2014ApJ...781...65W}.

Different from the nonthermal emission, as indicted in the soft X-ray band of $0.5-0.8 ~\mathrm{keV}$ \citep{2014ApJ...781...65W}, the thermal emission is distributed more uniformly inside the remnant; moreover, an obvious filament arises in the northwest (NW) quadrant of SN 1006, which is also indicated at the optical \citep{2002ApJ...572..888G,2007ApJ...659.1257R} and radio \citep{2009AJ....137.2956D} wavelengthes.

SN 1006 has a morphology in the hard X-rays with two opposite bright rims and protrusions and  a filament in the radio, optical and soft X-ray images. The formation mechanism of these peculiar structures is still in debate now. Assuming lower-density cavities located in the ambient environment of the ejecta, \citet{2015A&A...579A..35Y} explained the NE and SW bumps on the periphery of SN 1006 using 3D MHD simulations. Taking into account of a flat cloud parallel to the Galactic plane and different directions of the interstellar magnetic field, \citet{2010MNRAS.408..430S} obtained the synthetic radio and thermal X-ray emission maps for SN 1006 using 2D MHD axisymmetric simulations. This model explained the bright NW filament observed in optical wavelengths and in thermal X-rays, and the interstellar magnetic field perpendicular to the Galactic
plane was favored in the model. Assuming the remnant propagating in the turbulent interstellar medium, \citet{2017MNRAS.466.4851V} investigated the polarization properties of SN 1006. An adiabatic index of 1.3 for the turbulent medium was preferred, and the quasi-parallel case for the acceleration of the electrons could better reproduce the observed distribution of the Stokes parameter $Q$. For another remnant G1.9+0.1, which also has two opposite protrusions on the X-ray morphology, assuming it has evolved inside an elliptical planetary nebula with ears and clumps, \citet{2015MNRAS.450.1399T} investigated the formation of the shape with two ear-like protrusions using 3D HD simulations, and the interaction of the ejecta with the clumps produces the double-shock structure.

SNRs evolved in inhomogeneous mediums have morphologies with interesting signatures. In this paper, we intend to study the formation mechanism of the peculiar structures illustrated by the multiband observations for SN 1006. By introducing an ambient medium with a density discontinuity, we investigate how the substructures of it are produced via numerical simulations. Different from \citet{2010MNRAS.408..430S}, the HD simulations in this paper are performed in 3D, and the NE and SW protrusions and the NW filament can be reproduced using the model. The model and numerical setup for SN 1006 are shown In Section \ref{sect:model}. The results are presented in Section \ref{sect:result}. Finally, the discussion and some conclusions are given in Section \ref{sect:discon}.

\section{The model and numerical setup}
\label{sect:model}

\begin{figure}
        \centering
        \includegraphics[width=0.5\textwidth]{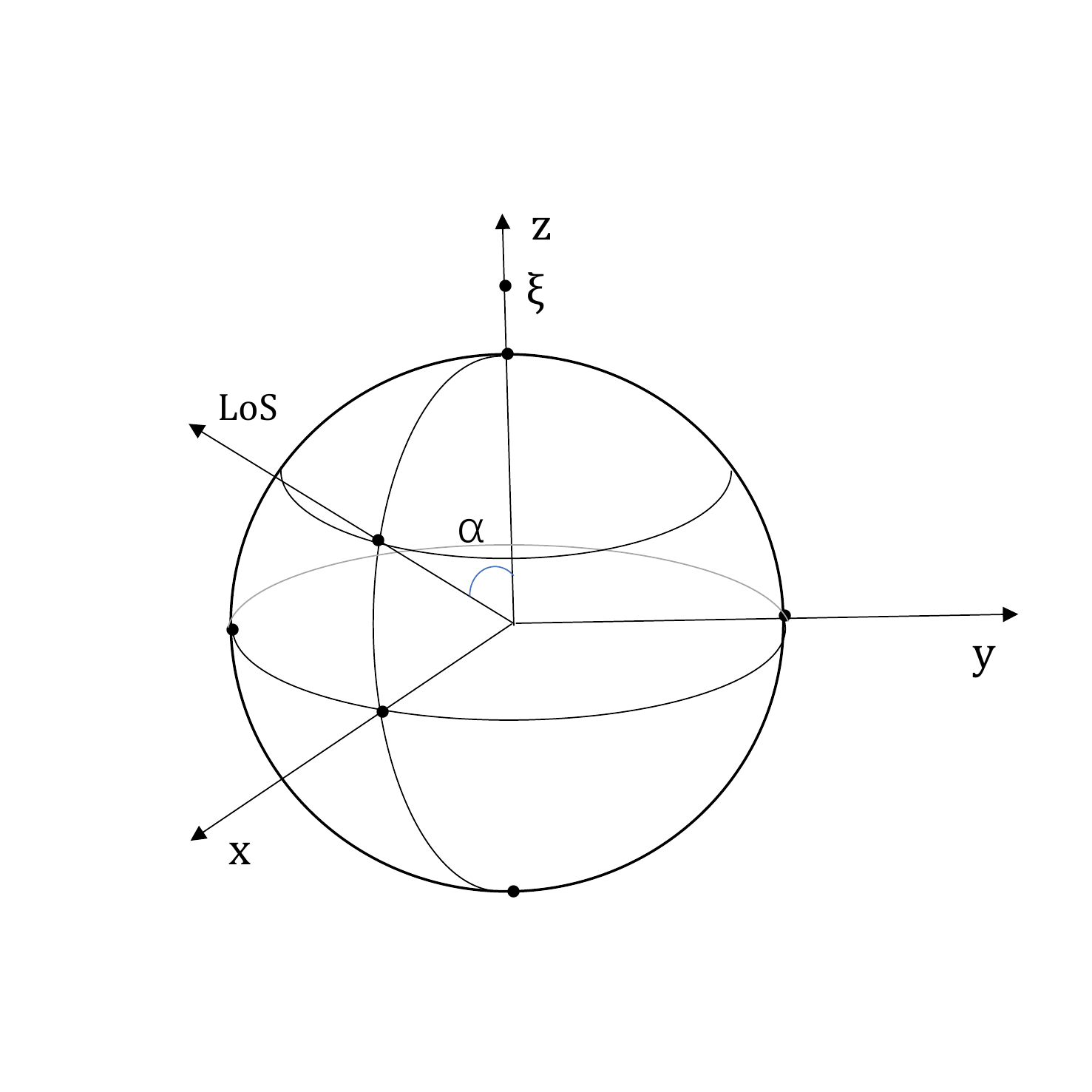}
        \caption{Scheme of the simulation model. The explosion occurs at the center, and there is a density discontinuity at the plane $\mathbf{r\cdot \hat{z}}=\xi$. The LoS is in the $xz$ plane with an angle of $\alpha$ to the $z$ axis.}
        \label{fig:model}
\end{figure}
To investigate the dynamical evolution of the remnant SN 1006 to reproduce the peculiar morphology, we assume the ejecta has evolved in the nonuniform medium with a density discontinuity. As illustrated in Fig.\ref{fig:model}, the plane of the density discontinuity is perpendicular to the $z$ axis, and the number density of the ambient medium is assumed to be
\begin{equation}
        n(\bf{r}) = \left\{  \begin{array}{cc}
    x n_{\mathrm a}  & \mathrm{if~} \bf{r}\cdot\bf{\hat{z}} > \xi \;, \\
    n_{\mathrm a} = \mu n_{\mathrm{H}} & \mathrm{otherwise} \;,
    \label{eq:nr}
   \end{array}
\right.,
\end{equation}
where $\hat{z}$ is the unit vector along the $z$ axis, $n_{\mathrm{H}}$ is the hydrogen number density, $\mu=1.4$ is the mean atomic mass for a gas of a 10 : 1 H:He ratio, the ratio $x=3$ is adopted in this paper, which is consistent with that derived from the observations with XMM-Newton \citep{2007A&A...475..883A}.

\begin{figure*}
        \centering
        \includegraphics[width=0.85\textwidth]{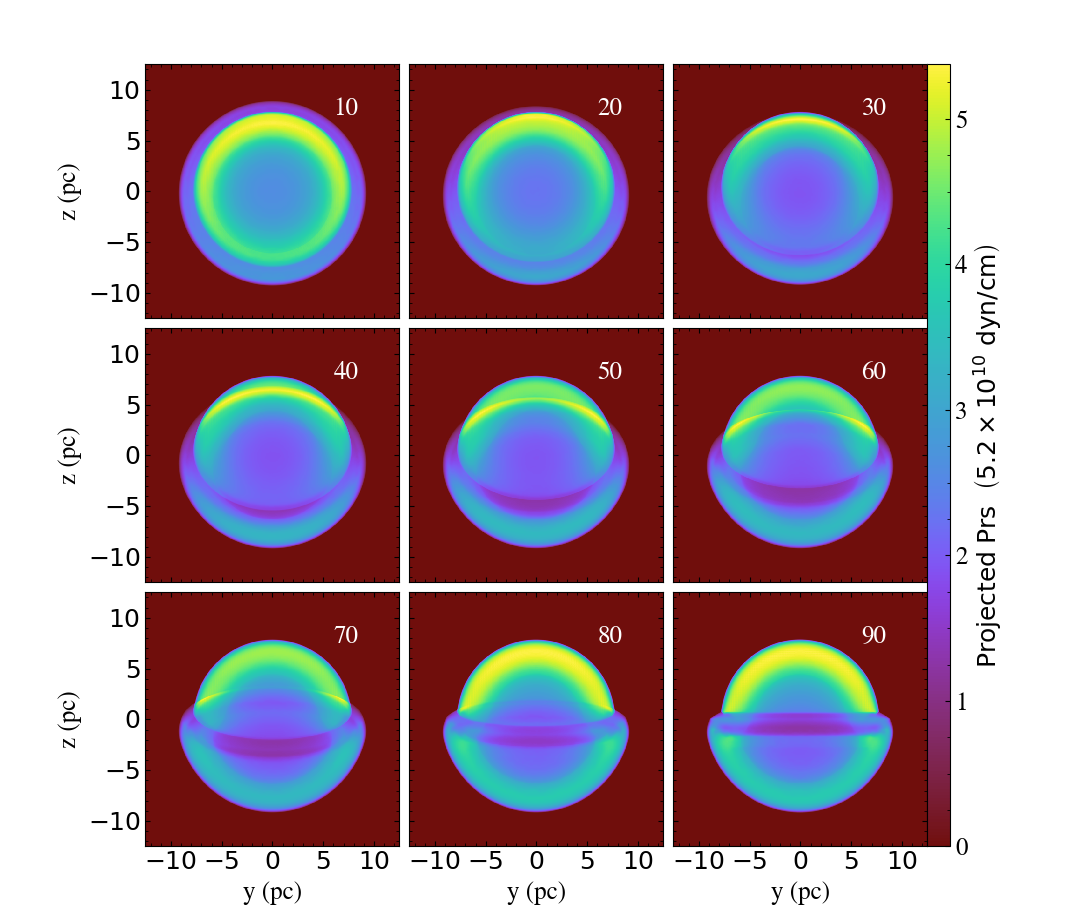}
        \caption{The morphologies of the projected pressure at $t=976\,\mathrm{yr}$ for different LoSs with $\alpha = 10^{\circ} - 90^{\circ}$. The parameters are $\xi=0$, $n_{\mathrm{a}}= 0.05\,\mathrm{cm}^{-3}$, $x=3.0$, $M_{\mathrm{ej}}=1.4M_{\odot}$, $E_{\mathrm{ej}}=10^{51}\,\mathrm{erg}$, $R_{\mathrm{ej}}=1.5\,\mathrm{pc}$.}
        \label{fig:sn1006los}
\end{figure*}

\begin{figure*}
        \centering
        \includegraphics[width=0.85\textwidth]{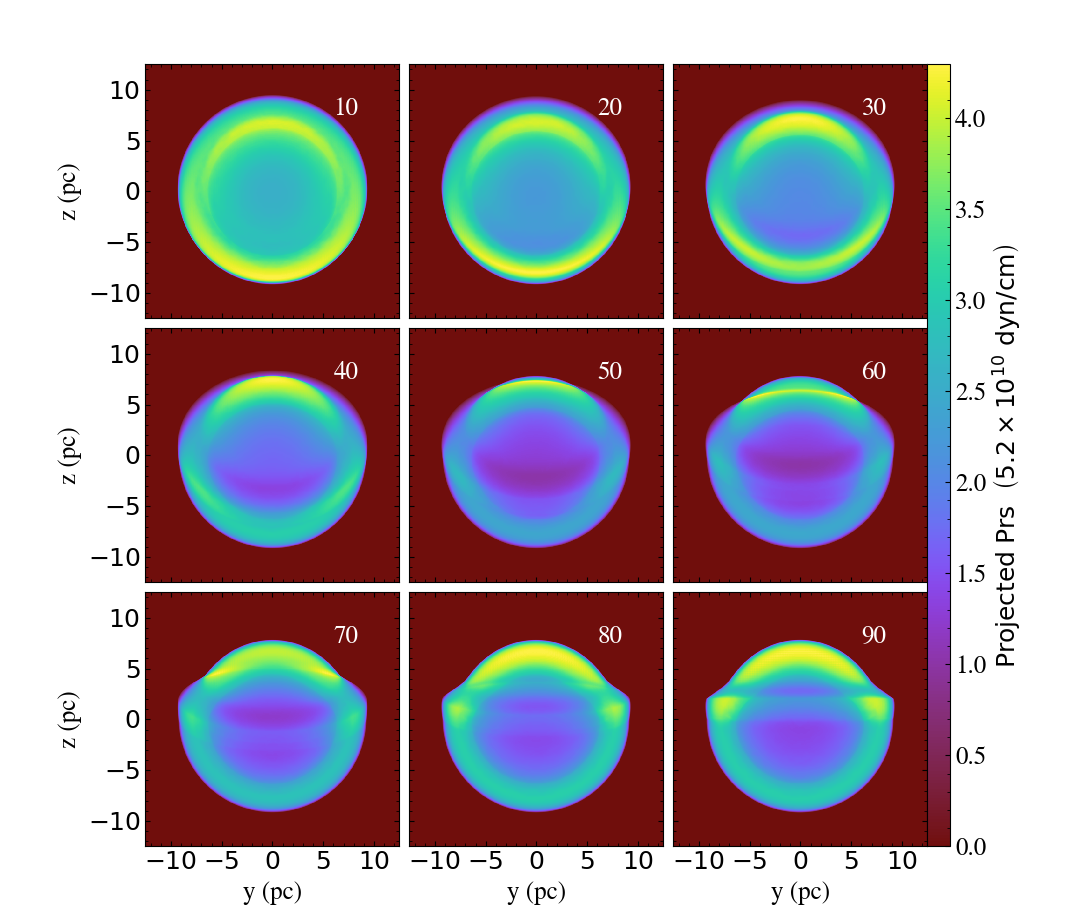}
        \caption{The morphologies of the projected pressure at $t=976\,\mathrm{yr}$ for different LoSs with $\xi=2\,\mathrm{pc}$. The other parameters are the same as Fig.\ref{fig:sn1006los}.}
        \label{fig:sn1006losksi2}
\end{figure*}

After the supernova explosion, the spherical ejecta with a radius of $R_{\mathrm{ej}}$, a mass of $M_{\mathrm{ej}}$ and a kinetic energy of $E_{\mathrm{ej}}$ is set at the center of the simulation. Initially, the material in the ejecta has a velocity of $v(\mathbf{r})= r/R_{\mathrm{ej}}v_0$, where $v_0$ is the velocity at the boundary of the ejecta \citep{1996ApJ...465..800J,1999ApJS..120..299T}. The inner part of the ejecta within a radius of $r_{\mathrm{c}}$ is homogenous with a density of $\rho_{\mathrm{c}}$, whereas the outer part has a mass of $\eta M_{\mathrm{ej}}$ and a density with a power-law dependence on $r$ with an index of $-s$ \citep{1996ApJ...465..800J}, i.e.,
\begin{equation}
\rho_{\rm ej}(r) = \left\{
  \begin{array}{cc}
    \rho_{\mathrm c}& \mathrm{if~} r < r_{\mathrm c}\;, \\
    \rho_{\mathrm 0}(r/R_{\mathrm ej})^{-s} & \mathrm{if~} r_{\mathrm c} < r <  R_{\mathrm{ej}}\;,
    \label{rho_c}
   \end{array}
\right.
\end{equation}
where $\rho_{\mathrm 0}$ is the density at $r=R_{\mathrm{ej}}$. $s=7$ and $\eta=3/7$ are used for the type Ia SNR in this paper. The other details of the initial condition for the ejecta were shown in the Eqs.1-3 in \citet{1996ApJ...465..800J}.

Without including the effects of the radiative cooling and the particle acceleration, the dynamical properties of the ejecta in the medium with a density discontinuity can be investigated using the Euler equations,
\begin{eqnarray}
\frac{\partial\rho}{\partial t} + \nabla\cdot(\rho \textbf{v}) & = & 0\; , \\
\frac{\partial \rho {\bf v}}{\partial t} + \nabla \cdot ( \rho {\bf
    vv} )   + \nabla{P} & = & 0\; , \\
\frac{\partial E}{\partial t} + \nabla \cdot (E+P){\bf v} ) & = & 0 ,
\end{eqnarray}
and
\begin{equation}
E = \frac{P}{\gamma - 1}+\frac{1}{2}\rho v^2 \;,
\nonumber
\end{equation}
where $P $ and $E$ are the gas pressure and the total energy density, respectively. $\gamma=5/3$ is adopted for the nonrelativistic gas. Using the PLUTO code \citep{2007ApJS..170..228M,2012ApJS..198....7M}, the simulations are performed in a cubic domain of $24\times24\times24$~pc$^{3}$ with equivalent $768\times768\times768$ grid cells for the remnant SN 1006.

\section{Results}
\label{sect:result}

The density of the medium around SN 1006 varies with azimuth. Based on the thermal emission detected with XMM-Newton from the NW and SE rims of the remnant, the density ranges from $\sim 0.05\, \mathrm{cm}^{-3}$ in the SE to $\sim 0.15 - 0.25 \,\mathrm{cm}^{-3}$ in the NW \citep{2007A&A...475..883A}. \citet{2014ApJ...781...65W} investigated the variation of the proper motions around the periphery of the remnant based on the observations with Chandra and the 4 m Blanco telescope at CTIO, and the velocity in the SE is about 2.5 times that in the NW. The lower expansion velocity in the NW is consistent with the higher-density ambient medium in this direction.

Initially, the density gradient is assumed to be along the $+z$ direction with $\xi=0$, $n_{\mathrm{snr}}= \mu n_{\mathrm{H}}=0.05\,\mathrm{cm}^{-3}$ and $x=3.0$, and the ejecta with $M_{\mathrm{ej}}=1.4M_{\odot}$, $E_{\mathrm{ej}}=10^{51}\,\mathrm{erg}$, $R_{\mathrm{ej}}=1.5\,\mathrm{pc}$ is set at the center of simulation. The material at the boundary of the ejecta has a high velocity of $v_0= 2 \times 10^9 \, \mathrm{cm}\,\mathrm{s}^{-1}$, and it drives a shock to heat the ambient medium. The shocked ambient medium drives a reverse shock heating the ejecta, and Rayleigh–Taylor instabilities arise around the contact discontinuity which separates the shocked ambient medium and the shocked ejecta \citep{2007A&A...470..927O,2018MNRAS.474.2544F}.  In the region of the instabilities, the magnetic field is amplified with a turbulent structure, and the synchrotron emission can be enhanced if relativistic electrons are distributed in this area. Moreover, the Rayleigh–Taylor features can also be indicated in the image of the thermal X-ray emission \citep{2010MNRAS.408..430S}. The emissions close to the instabilities are usually well inside the shell of the remnant \citep{2007A&A...470..927O,2010MNRAS.408..430S}. 

Fig.\ref{fig:sn1006los} shows the morphology of the projected pressure, i.e., $\int P \mathrm{d}l$, where $\mathrm{d}l$ is the increment along the line of sight (LoS), for different LoSs, whose directions are in the $xz$ plane with various angles ($\alpha$) to the $+z$ direction. In the dense medium, the forward shock expands more slowly with an average radius of $\sim 7.5\,\mathrm{pc}$ at $t=976 \mathrm{yr}$, whereas it is $\sim 9 \,\mathrm{pc}$ for the forward shock in the tenuous material. For the LoSs with $\alpha$ smaller than $\sim20^{\circ}$, the morphology of the projected pressure shows a double-shock structure since the edge of the hemisphere in the dense medium is well surrounded by that with a larger radius. Between $30^{\circ}$ and $70^{\circ}$, the periphery is of a distorted oval with a bump in the north, and a filament arises due to the intersection of the LoS and the shell of the two hemispheres. Finally, the morphology consists of two semicircular shells with different radius for $\alpha \sim 90^{\circ}$.

The influence of $\xi$ on the resulting morphology of the projected pressure is indicated in Fig.\ref{fig:sn1006losksi2} with $\xi=2\,\mathrm{pc}$.  With $\alpha \leq 40^{\circ}$, the shock ranging from the small hemisphere is well inside that from the large one. For $\alpha = 50^{\circ}-70^{\circ}$, a filament is produced due to the intersection of the LoS and the materials near the edges of the two hemispheres. The hemisphere in the dense medium has a smaller extension, and the filament for $\alpha=50^{\circ}$ also has a smaller azimuth extension compared with $\xi=0$. Finally, a belt is prominent in the middle of the images for $\alpha = 80^{\circ}$ and $90^{\circ}$.

\begin{figure}
        \centering
        \includegraphics[width=0.35\textwidth]{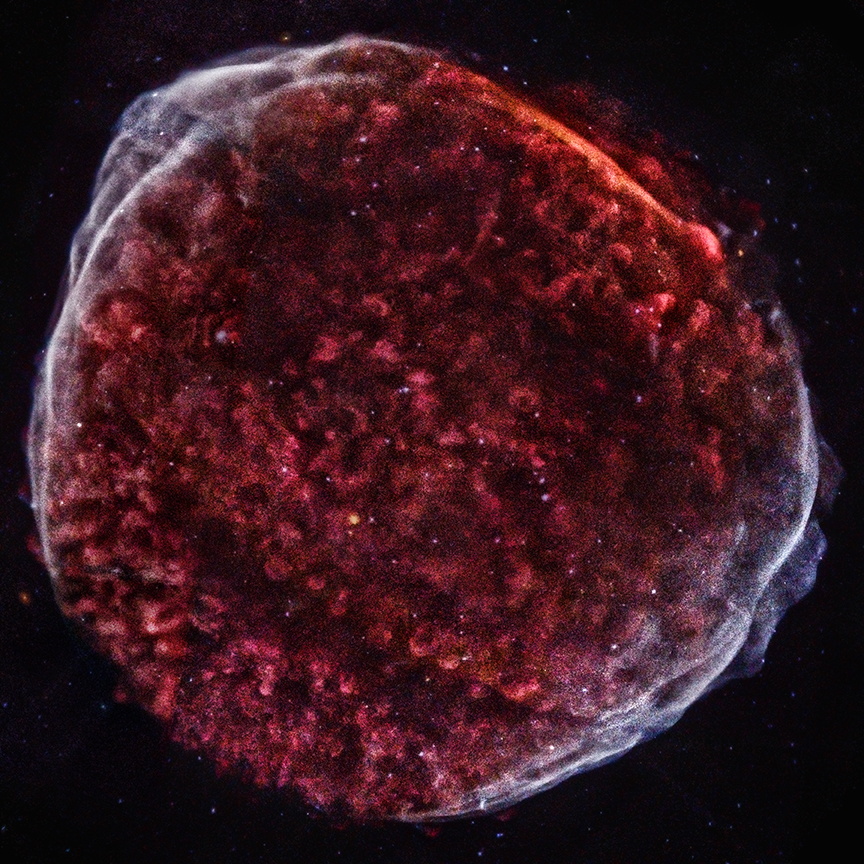}
        \caption{Upper panel: Chandra image of SN 1006 in the soft, medium, hard X-rays shown in red, green, blue colors, respectively (Credit: NASA/CXC/Middlebury College/F.Winkler). This image is downloaded from the website https://chandra.harvard.edu/photo/2013/sn1006/.}
        \label{fig:sn1006chan}
\end{figure}

\begin{figure}
        \centering
        \includegraphics[width=0.5\textwidth]{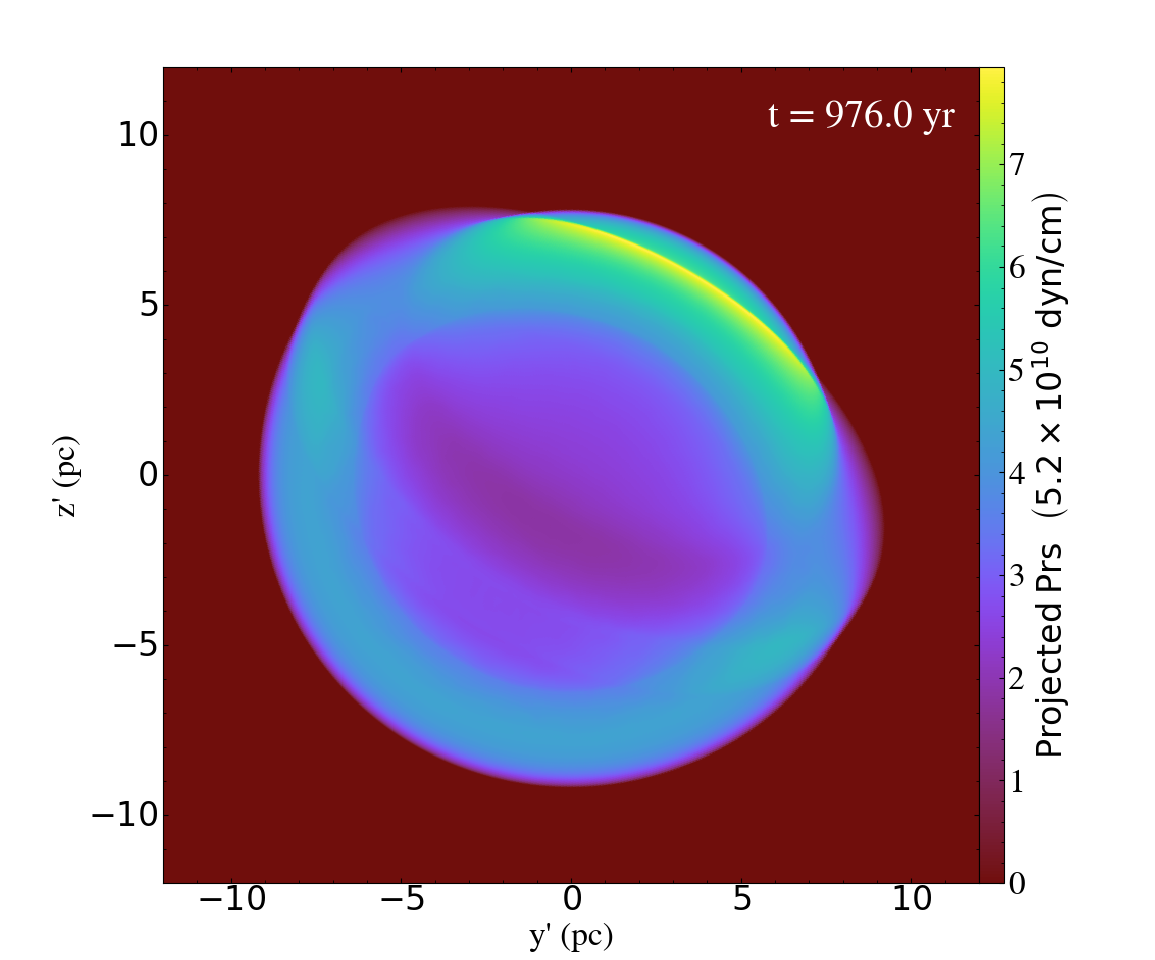}
        \caption{The morphology of the projected pressure, which is $30^{\circ}$ rotated clockwise in the plane perpendicular to the LoS for comparison with the observed X-ray image, at $t=976\,\mathrm{yr}$ with $\xi=2\,\mathrm{pc}$, $\alpha = 55^{\circ}$. The other parameters are the same as Fig.\ref{fig:sn1006losksi2}.}
        \label{fig:sn1006model}
\end{figure}

As illustrated in the image obtained with Chandra (Fig.\ref{fig:sn1006chan}), SN 1006 has a bilateral morphology in the hard X-rays with the bright NE and SW limbs and prominent protrusions; moreover, in the soft X-rays, a filament arises in the NW of the remnant, and some materials at the NW boundary  beyond the filament emit soft X-rays. Fig.\ref{fig:sn1006model} shows the morphology of the projected pressure in the model of $\xi=2\,\mathrm{pc}$ for the LoS with $\alpha = 55^{\circ}$, which is $30^{\circ}$ rotated clockwise in the plane perpendicular to the LoS for comparison with the detected X-ray image.  Protrusions in the NE and SW of the periphery can be produced because the shocked material around the boundary of the hemisphere evolved in the tenuous medium is located outside the shell of the small hemisphere in the image. Furthermore, the detected magnetic field around the remnant is quasi-parallel for the shock in the NE and SW \citep{2013AJ....145..104R}. Electrons can be effectively accelerated in these regions via the diffusive shock acceleration, and two bright lobes with protrusions in the hard X-rays are formed. In the NW, a filament arises, and some shocked material from the hemisphere evolved in the dense medium is located beyond the filament in the image of the projected pressure. These features on the protrusions in the NE and SW and a filament in the NW are consistent with those derived from the detected X-ray image.

\section{Discussion and conclusions}
\label{sect:discon}

Multiband observations on SNRs show their morphologies vary with each other. The dynamical evolution and the morphology of a SNR depend heavily on the anisotropy and inhomogeneity of both the ejecta and the ambient medium. In this paper, we study the dynamics of the initially isotropic ejecta within the medium with a density discontinuity. The shell consists of two hemispheres with different radius, and the resulting morphology of the projected pressure varies with the LoS. Moreover, some peculiar features illustrated in the detected images, such as protrusions on the periphery and a filament, can be reproduced with the model. In this paper, the morphology of the projected pressure of the remnant along a LoS is used to show the protrusions and the filament, and the resulting morphology of emission is not investigated because more assumptions on the distribution of the high-energy electrons and the magnetic field in the remnant are necessary to obtain it.

The model in which the ambient medium has a density discontinuity is applied to the type Ia SNR SN 1006 to study how the peculiar properties of the morphology are formed.  The model is intrinsically axisymmetric, and the dynamical evolution of the remnant can be solved with 2D axisymmetric HD simulations. However, the simulations are performed in 3D in this paper because the morphology indicated by the projected pressure along different LoSs can be easily obtained. Moreover, we pay attention to the protrusions on the NE and SW limbs and the NW filament indicated in the radio and X-ray images, which are illustrated in the image of the projected pressure even without the morphology of the synchrotron emission. The images of the remnant in the hard X-rays shows prominent protrusions on the NE and SW limbs. Our results show that these protrusions  can be explained as the emission of the materials from the hemisphere with a larger radius, and the emission seems to be located outside the boundary of the hemisphere with a smaller radius for certain LoSs. The NW filament of SN 1006 arises as a result of the intersection of LoS and the shells in the two hemispheres. \citet{2015A&A...579A..35Y} explained the protrusions in the X-ray image of SN 1006 by assuming low-density cavities around the ejecta, which differs significantly from our model.

Based on the detected morphology of SN 1006 in the X-rays as indicated in Fig.\ref{fig:sn1006chan}, the protrusions in the NE and SW of the remnant are not symmetrical, and there is an obvious hollow in the SW protrusion. This hollow can be interpreted as the SW limb interacting with a uniform cloud \citep{2014ApJ...782L..33M,2016A&A...593A..26M}, which is not included in this paper because we focus primarily on the formation of the features of the protrusions on the NE and SW limbs and the filament in the NW. \citet{2010MNRAS.408..430S} employed 2D MHD simulations to reproduce the filament by assuming a flat cloud in the NW of the remnant. Our simulations are performed in 3D, and the NE and SW protrusions and the NW filament can be simultaneously reproduced.

\section*{Acknowledgements}
JF is partially supported by the National Key R\&D Program of China under grant No.2018YFA0404204, the
Natural Science Foundation of China (NSFC) through grants 11873042 and 11563009,
the Yunnan Applied Basic Research
Projects under (2018FY001(-003)), the Candidate Talents Training Fund of Yunnan Province (2017HB003)
and the Program for Excellent Young Talents, Yunnan University (WX069051, 2017YDYQ01).
\setlength{\bibhang}{2.0em}
\setlength\labelwidth{0.0em}
\bibliography{snl006mn}

\end{document}